\documentclass[12pt,preprint]{aastex}

\def\be{\begin{equation}}
\def\ee{\end{equation}}

\catcode`\@=11 
\def\@versim#1#2{\vcenter{\offinterlineskip
        \ialign{$\m@th#1\hfil##\hfil$\crcr#2\crcr\sim\crcr } }}

\def\rms{r_{ms}}

\shorttitle{****}
\shortauthors{****}
\begin{document}

\title{A simplified global solution for an advection-dominated accretion flow}

\author{Feng Yuan\altaffilmark{1,2}, Renyi Ma\altaffilmark{1},
Ramesh Narayan\altaffilmark{3}}
\altaffiltext{1}{Shanghai Astronomical Observatory, Chinese Academy of Sciences,
80 Nandan Road, Shanghai 200030, China; fyuan@shao.ac.cn}
\altaffiltext{2}{Joint Institute for Galaxy and Cosmology (JOINGC) of 
SHAO and USTC, China}
\altaffiltext{3}{Harvard-Smithsonian Center for Astrophysics, 60 Garden Street,
Cambridge, MA 02138, U.S.A.}

\begin{abstract}

When we model black hole accretion sources such as active galactic
nuclei and black hole X-ray binaries as advection-dominated accretion
flows (ADAFs), it is neccesary to use the global solution to the
equations rather than the simpler self-similar solution, since the
latter is inaccurate in the region near the black hole where most of
the radiation is emitted. However, technically, it is a difficult task
to calculate the global solution because of the transonic nature of
the flow, which makes it a two-point boundary value problem.  In this
paper we propose a simplified approach for calculating the global ADAF
solution.  We replace the radial momentum equation by a simple
algebraic relation between the angular velocity of the gas and the
Keplerian angular velocity, while keeping all other equations
unchanged. It is then easy to solve the differential energy equations
to obtain an approximate global solution. By adjusting the free
parameters, we find that for almost any accretion rate and for
$\alpha=0.1-0.3$ we can get good simplified global solutions.  The
predicted spectra from the approximate solutions are very close to the
spectra obtained from the true global solutions.

\end{abstract}


\section{Introduction}
\label{intro}

Advection-dominated accretion flow (ADAF) is an important type of
solution for black hole accretion. A prominent feature of an ADAF
compared to the standard thin disk is its low radiative efficient at
low accretion rates (Narayan \& Yi 1994, hereafter NY94; Narayan \& Yi
1995; see Narayan, Mahadevan \& Quataert 1998 and Kato, Fukue \&
Mineshige 1998 for reviews). 

The ADAF solution has received much attention in the past years
because it successfully explains why some nearby galaxies are so dim
even though their accretion rates are not very small (see Narayan
2005, Yuan 2007, and Ho 2008 for reviews). The best evidence comes
from the supermassive black hole in our Galactic center, Sgr A*
(Narayan, Mahadevan \& Yi 1995; Manmoto, Kusunoze \& Mineshige 1997;
Yuan, Quataert \& Narayan 2003). From {\it Chandra} observations and
Bondi theory, we can estimate the mass accretion rate of Sgr A*.  If
the accretion flow were not an ADAF but a standard thin disk, the
luminosity would be five orders of magnitude larger than observed
(Yuan, Quataert \& Narayan 2003).

Another attractive feature of an ADAF is that it can partly solve the
problem of the origin of X-ray emission from accretion flows.  The
temperature of a standard thin disk at the inner disk is only $\sim
10^{5}$K for a supermassive black hole, too low to produce X-ray
emission (Frank, King \& Raine 2002). A hot corona has been thought to
be responsible for the X-ray emission, but recent MHD simulations of
disks show that they have hardly any coronae (Hirose, Krolik \& Stone
2006).  On the other hand, the temperature of an ADAF is high enough
to produce X-ray emission. Of course, a canonical ADAF exists only
below a critical accretion rate $\dot{M}_{\rm crit} \approx
\alpha^2\dot{M}_{\rm Edd}$ with $\dot{M}_{\rm Edd}\equiv 10L_{\rm
Edd}/c^2$ and $\alpha$ is the viscous parameter, which corresponds to
$\sim (3-4)\%L_{\rm Edd}$ at most (Esin, McClintock \& Narayan
1997). Therefore it cannot explain luminous X-ray sources such as
quasars. The luminous hot accretion flow (LHAF; Yuan 2001), which is
an extension of an ADAF to higher accreton rates, is promising, but
the details of this model have not been fully worked out (see Yuan et
al. 2007 for an example of application to luminous black hole X-ray
binaries).

In spite of the great success of ADAFs, more work is required to test
the model. On the one hand, it would useful to expand the application
of ADAFs to more sources, and on the another hand, such modeling is
expected to help us understand some important microphysical issues
which are still unclear. One example of the latter is the value of
$\delta$ (defined in eqs. 4 and 5), which measures the amount of
direct electron heating through viscous dissipation in a hot accretion
flow (Sharma et al. 2007).  Another is the potential importance of
collective plasma effects which will determine how realistic the
two-temperature assumption is (Begelman \& Chiueh 1988; Yuan et
al. 2006).

For such work, the global solution rather than the self-similar
solution of the ADAF equations is required. This is because most of
the radiation of an ADAF comes from its innermost region where the
self-similar solution breaks down. However, it is technically very
difficult to calculate the global solution of an ADAF.  An ADAF is
transonic, and thus its global solution should satisfy the sonic-point
condition in addition to the outer boundary condition. Mathematically,
it is a two point boundary value problem and not easy to deal
with. This is an obstacle to the wide application of the ADAF model.

In this paper we propose a simplified global ADAF solution.  We adopt
a simple algebraic relation to replace the radial momentum equation,
thus avoiding the two point boundary value problem. We present our
approach in \S 2 and show some examples in \S 3.  The final section is
devoted to a short summary.  Watarai (2007) has recently presented
related work.

\section{The simplified global ADAF model}
 
The basic equations of an ADAF describe the conservation of mass,
radial and aximuthal components of the momentum, and energy (e.g.,
Narayan, Mahadevan \& Quataert 1998):

\begin{equation}
\dot{M}=-4\pi r H \rho v
\end{equation}
\begin{equation}
v\frac{dv}{dr}=(\Omega^2-\Omega_K^2)r-\frac{1}{\rho}\frac{dP}{dr}
\end{equation}
\begin{equation}
v(\Omega r^2-j)=-\alpha r c_s^2
\end{equation}
\begin{equation}
\rho v T_i\frac{ds_i}{dr}=(1-\delta) q^+-q_{ie}
\end{equation}
\begin{equation}
\rho v T_e\frac{ds_e}{dr}=q_{ie}-q^--\delta q^+
\end{equation}
All the quantities have their usual meaning. In the present paper we
do not include outflows from the ADAF, but it is easy to extend our
calculation to that case by simply using a radius-dependent mass
accretion rate, $\dot{M}=\dot{M}_0(r/r_{\rm out})^s$ with $s>0$ being
a constant (e.g., Yuan, Quataert \& Narayan 2003).  The quantity
$\delta$ in equations (4) and (5) describes the fraction of the
turbulent dissipation rate $q^+$ which directly heats electrons; we
set $\delta=0.3$.  The quantity $q_{ie}$ describes the energy transfer
rate from ions to electrons by Coulomb collision, and $q^-$ is the
radiative cooling rate. We consider synchrotron and bremsstrahlung
emissions and their Comptonization. The details of the calculation of
the spectrum can be found in Yuan, Quataert \& Narayan (2003).  We
consider a Schwarzschild black hole and adopt the Paczy\'nski \& Wiita
(1980) potential to mimic its geometry.

The most difficult part of 
solving the global solution is the radial momentum equation (2).
Our key idea of simplifying the global ADAF solution is to
replace this differential equation by the following simple algebraic relation:
\begin{equation}
\Omega=f \Omega_K,
\end{equation}
with
\[f=\left\{ \begin{array}{l@{\quad {\rm for} \quad}l}
f_0  &  r>\rms \\ \frac{f_0\Omega_K(\rms)\rms^2}{\Omega_Kr^2}\left(\frac{r}
{\rms}\right)^n=f_0\frac{r-r_g}{2r_g}(\frac{r}{\rms})^{n-3/2} & r<\rms.
\end{array} \right.\]
where $\rms=3r_g\equiv6GM/c^2$ is the innermost stable circular orbit.

The above simplification is based on the following physical
consideration.  The immediate idea we think of to simplify the radial
momentum equation is to use the self-similar solution obtained by
NY94.  Consider eqs. (7)-(9) in NY94.  We can use eq. (9) to solve for
$\epsilon^{\prime}$ in terms of the sound speed $c_s$: \be
5+2\epsilon^{\prime}=2\frac{v_K^2}{c_s^2}, \hspace{1cm}
\epsilon^{\prime}=\frac{v_K^2}{c_s^2}-\frac{5}{2}.  \ee We can then
substitute this in eqs. (8) and (7) in NY94 to obtain for a fully
advection-dominated flow ($\epsilon^{\prime}=\epsilon$) \be
\frac{\Omega}{\Omega_K}=\left(1-\frac{5c_s^2}{2v_K^2}\right)^{1/2}
=\frac{10-6\gamma}{9\gamma-5}={\rm const}.  \ee Here $\gamma$ is the
adiabatic index.  We therefore in principle could set $f$ in eq. (6)
to this constant.  However, we find that the simplfied solution is
very sensitive to the value of $f$. The reason is that, as we will
see, the radial velocity is sensitive to the value of $f$ (ref. eq. 9
below). The velocity determines the density, and also the temperature
via the energy equations, two quantities that determine the emitted
spectrum.  We therefore set $f$ as a free parameter which we adjust
for different accretion parameters $\dot{M}$ and $\alpha$ to get the
best approximation.

Because the angular momentum $\Omega r^2$ in a global solution keeps
decreasing with decreasing radius, while the Keplerian angular
momentum $\Omega_K r^2$ begins to increase when $r<\rms$ (Fig. 1), $f$
cannot be a constant when $r<\rms$.  Instead we require the angular
momentum to be continuous at $\rms$ and assume that it is proportional
to $(r/\rms)^n$. After some tests we set $n=0.5$, independent of the
values of $\dot{M}$ and $\alpha$. Thus $n$ is not a free parameter in
our model.

Substituting eq.(6) into eq.(3) we have
\begin{equation}
v_r=-\frac{\alpha r c_s^2}{f\Omega_K r^2-j} .
\end{equation}
The quantity $j$ is the specific angular momentum of the accretion gas
when it falls into the black hole and it is the eigenvalue of 
the exact global solution. In our simplified model, 
we set $j$ as the second free parameter and adjust its value to get 
the best approximate solution for $v$.

Substituting eqs. (1), (6) and (9) into the energy equations for ions
and electrons, eqs. (4) and (5), we have two differential equations
with two unknown variables, $T_i$ and $T_e$.  All other quantites such
as $v$, $\rho$, $c_s$ and $H$ can be expressed as simple functions of
$T_i$ and $T_e$ for a given $\dot{M}$ and $\alpha$ and assumed values
of the free parameters $f_0$ and $j$. When $T_i$ and $T_e$ are given
at the outer boundary, we can easily integrate the differential
equations inwards to get the approximate global solution.

We adjust the values of $f_0$ and $j$ for different $\dot{M}$ and
$\alpha$ to get the best simplified global solution. Here ``best''
means that the profiles of all quantities such as $\rho, ~v, ~T_e,
~T_i$, and most importantly, the emitted spectrum, are very close to
the exact global solution.  Because our main purpose is to model the
continuum spectrum of black hole sources (AGNs and black hole X-ray
binaries), our first priority will be the closeness of the spectrum
when we judge how good a simplified solution is. 

In the calculations presented here, we set $\dot{M}$ of the
approximate solution equal to $\dot{M}$ of the global solution it is
meant to fit.  However, in real applications, we only know the
spectrum rather than $\dot{M}$.  So it might be more realistic to
adjust $\dot{M}_{\rm simp}$ of the simplified solution to fit the
spectrum produced by the exact global solution with a given
$\dot{M}_{\rm exact}$ rather than setting $\dot{M}_{\rm
simp}=\dot{M}_{\rm exact}$. Fortunately we find that $\dot{M}_{\rm
simp}$ and $\dot{M}_{\rm exact}$ are very close, typically eviating by
no more than $\sim 3\%$.

\section{Results}

When modeling black hole sources with an ADAF model, the accretion
rate $\dot{M}$ spans a wide range, say from $10^{-6} \dot{M}_{\rm
Edd}$ to $10^{-1}\dot{M}_{\rm Edd}$. But the value of $\alpha$ adopted
in ADAF modeling (e.g., Narayan, Mahadevan \& Quataert 1998) is
usually within a very narrow range, $\alpha=0.1-0.3$. This is also
supported by MHD numerical simulations of accretion flows (e.g.,
Hawley \& Krolik 2001). We adjust the values of $j$ and $f_0$ to
obtain the ``best'' simplified ADAF solutions for $\dot{M}$ and
$\alpha$ within the above ranges. As state below we find that the same
set of ($j, f_0$) often holds for quite a wide range of $\dot{M}$.

\subsection{$\alpha=0.3$: $f_0=0.33, ~j=0.98$ for any $\dot{M}$}

We first present results for $\alpha=0.3$. We find that in this case
the simplified global solution with $f_0=0.33, ~j=0.98$ gives a
satisfactory spectrum for any $\dot{M}$. Figs. 1 \& 2 give two
examples with $\dot{M}=10^{-5}$ and $10^{-1} \dot{M}_{\rm Edd}$,
respectively. The dashed lines in the figure denote the exact global
solution while the solid lines are for the simplified global
solution. The plots in each figure show the emitted spectrum, Mach
number, electron and ion temperature, density, and the angular
momentum, respectively. For Fig. 1, the outer boundary is at $10^4r_g$
and the outer boundary condition is $T_i=0.2T_{\rm vir}, ~T_e=0.19
T_{\rm vir}$ with the virial temperature $T_{\rm vir}\equiv 3.6\times
10^{12}(r_g/r)$. For Fig. 2, the outer boundary is at $10^2r_g$ and
the outer boundary condition is $T_i=0.6T_{\rm vir}, ~T_e=0.08 T_{\rm
vir}$.

\subsection{$\alpha=0.1$: $f_0=0.33, ~j=1.08$ for 
$\dot{M}\la 10^{-2}\dot{M}_{\rm Edd}$}

When $\alpha=0.1$, it is hard for a single set of ($j,f_0$) to give a
good solution for all $\dot{M}$. When $\dot{M}$ is relatively low,
$\dot{M}\la 10^{-2}\dot{M}_{\rm Edd}$, we find $f_0=0.33, ~j=1.08$
gives a satisfactory solution. Fig. 3 shows an example with
$\dot{M}=10^{-3}\dot{M}_{\rm Edd}$. The outer boundary conditions are
$T_i=0.2 T_{\rm vir}, ~T_e=0.19T_{\rm vir}$ at $r_{\rm out}=10^4r_g$.

\subsection{$\alpha=0.1$: $f_0=0.15, ~j=0.49$ for $\dot{M}\ga 5\times 
10^{-2}\dot{M}_{\rm Edd}$}

When $\dot{M}$ is relatively high, $\dot{M}\ga 5\times
10^{-2}\dot{M}_{\rm Edd}$, we find $f_0=0.15, ~j=0.49$ gives a
satisfactory solution. Fig. 4 shows an example of $\dot{M}=10^{-1}
\dot{M}_{\rm Edd}$. The outer boundary conditions are $T_i=0.6T_{\rm
vir}, ~T_e=0.08T_{\rm vir}$ at $r_{\rm out}=10^2r_g$.  Note this
solution is in the regime of LHAF, because $10^{-1}\dot{M}_{\rm Edd}$
is well above the critical accretion rate of an ADAF $\dot{M}_{\rm
crit}\approx 10^{-2}\dot{M}_{\rm Edd}$.

\subsection{Other values of $\alpha$}

For other values of $\alpha$, we find that simply using 
the ``linear fit'' values of $(j,f_0)$ between those for $\alpha=0.3$ 
and $0.1$ gives a good solution. For example, 
for $\alpha=0.2$ and $\dot{M}=10^{-3}\dot{M}_{\rm Edd}$, 
the values of $j$ and $f_0$ are just $f_0=(0.33+0.33)/2=0.33, ~j=(0.98+
1.08)/2=1.03$. For $\alpha=0.2$ and $\dot{M}=0.08\dot{M}_{\rm Edd}$,
$f_0=(0.33+0.15)/2=0.24, ~j=(0.98+0.49)/2=0.735$.

\section{Summary}

The global solution of ADAFs is difficult to calculate because it is
mathematically a two point boundary value problem. This hampers wide
application of the ADAF model. We propose a simplifed global solution
to overcome this difficulty.  Prompted by the self-similar solution of
ADAFs, we replace the radial momentum equation, which is the most
difficult to handle, with a simple algebraic relation (eq. 6), and
then solve the remaining two diferential equations (eqs. 4 \& 5). We
adjust the two free parameters ($j$ and $f_0$ in eqs. 6 and 9) so that
we obtain the best approximation compared to the exact global ADAF
solution. The spectra of the simplified solutions are impressively
good, as shown in Figs. 1-4 for various $\dot{M}$ and $\alpha$.

We have been unable to identify a single set of values of the two
adjustable constants $j$ and $f_0$ which works for all ADAF models.
However, when $\alpha$ is large, say $\sim0.3$, we find that $j=0.98$,
$f_0=0.33$ gives very good results for all accretion rates $\dot{M}$
for which an ADAF solution is possible.  Recent work by Sharma et
al. (2006) suggests that the viscosity parameter in the collisionless
plasma in a hot accretion flow will be larger than in a standard thin
disk.  Therefore, $\alpha\sim0.3$ is probably not unrealistic for an
ADAF.  It would thus be reasonable to use a single set of parameters,
$\alpha=0.3$, $j=0.98$, $f_0=0.33$, for practical applications of the
approximate global model described here.

The success of the present work encourages us to extend our approach
to the case of a slim disk, which is an extension of the standard thin
disk to accretion rates above the Eddington rate (Abramowicz et
al. 1988).  It potentially has important application in ULXs and
narrow-line Seyfert 1 galaxies (Mineshige et al. 2000; Watarai et
al. 2001). We hope to report the results in a future paper.
 
\acknowledgements

This work was supported in part by the Natural Science Foundation of China 
(grant 10773024), Shanghai Pujiang Program, Bairen Program of CAS (F.Y.), 
the Knowledge Innovation Program of CAS, and 
Shanghai Postdoctoral Scientific Program (R.M.).

{}

\clearpage

\begin{figure} \epsscale{0.9} \plotone{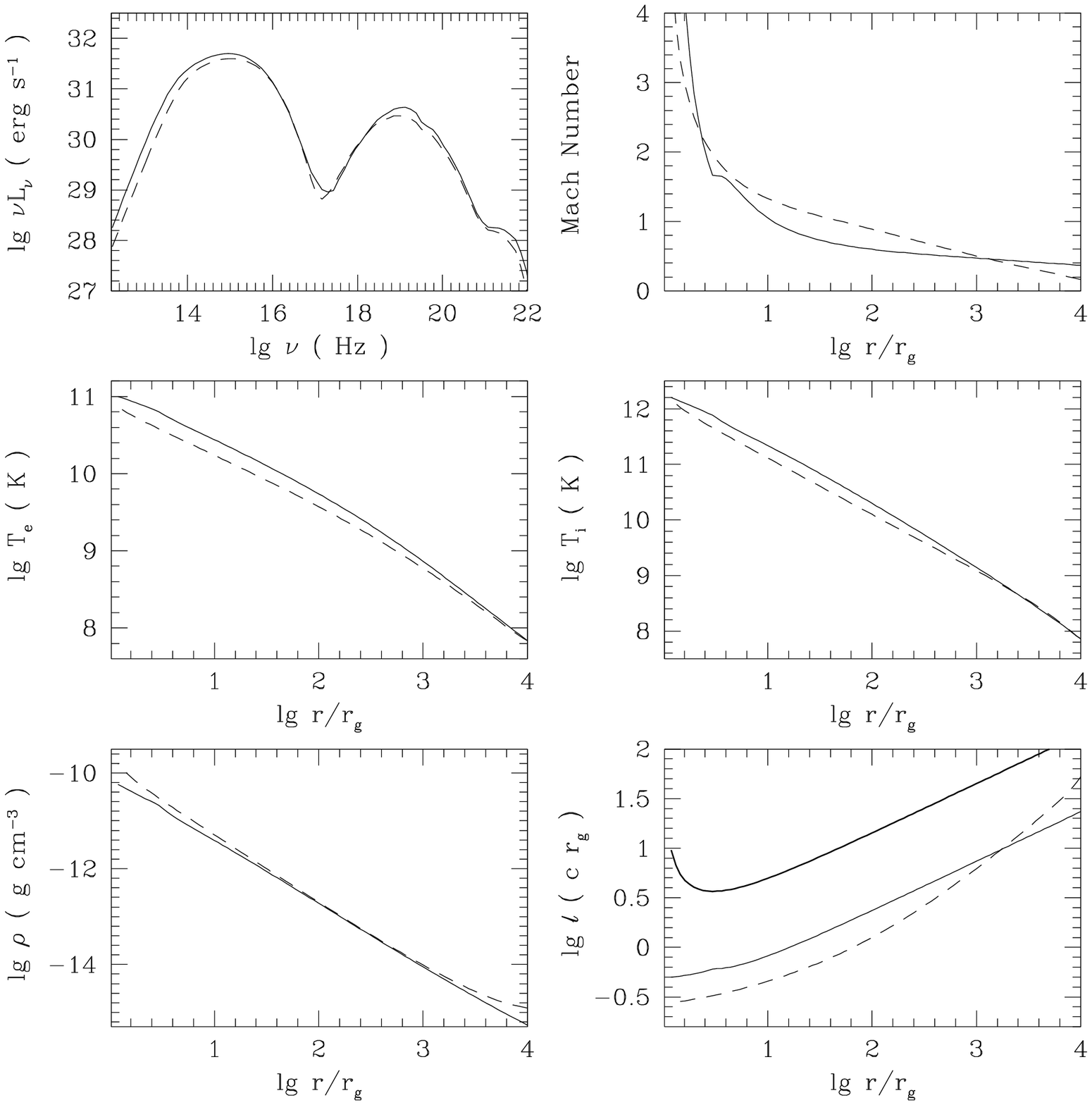} \vspace{.2in}
\caption{The simplified (solid) and exact (dashed) global ADAF solutions
for $\dot{M}= 10^{-5}\dot{M}_{\rm Edd}, ~\alpha=0.3$. The 
parameters are $f_0=0.33, ~j=0.98$.
\vspace{.4in}
\label{spectra}} 
\end{figure}

\begin{figure} \epsscale{0.9} \plotone{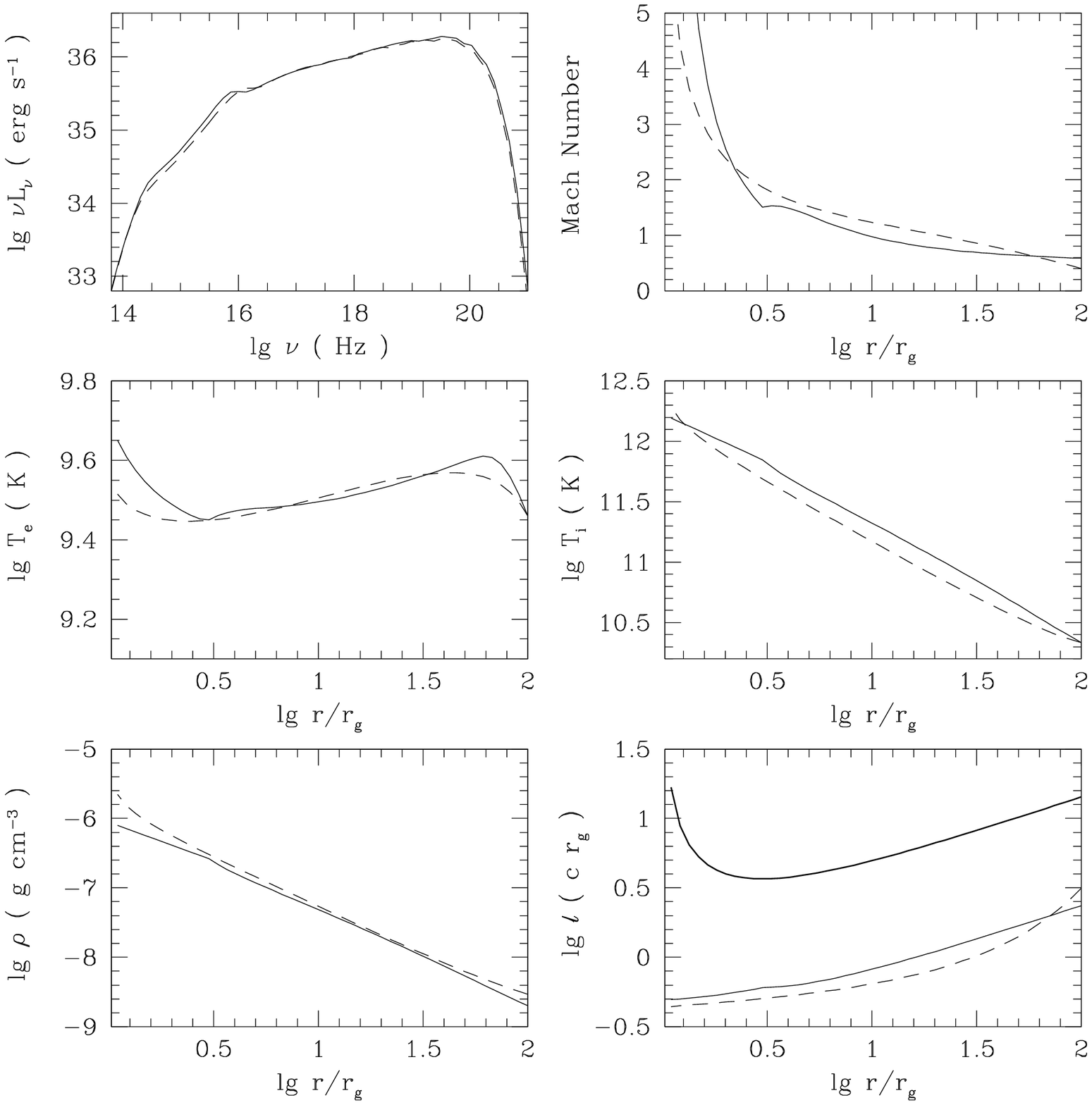} \vspace{.40in}
\caption{The simplified (solid) and exact (dashed) global ADAF solutions
for $\dot{M}= 10^{-1}\dot{M}_{\rm Edd}, ~\alpha=0.3$. The
parameters are $f_0=0.33, ~j=0.98$.}
\end{figure}

\begin{figure} \epsscale{0.9} \plotone{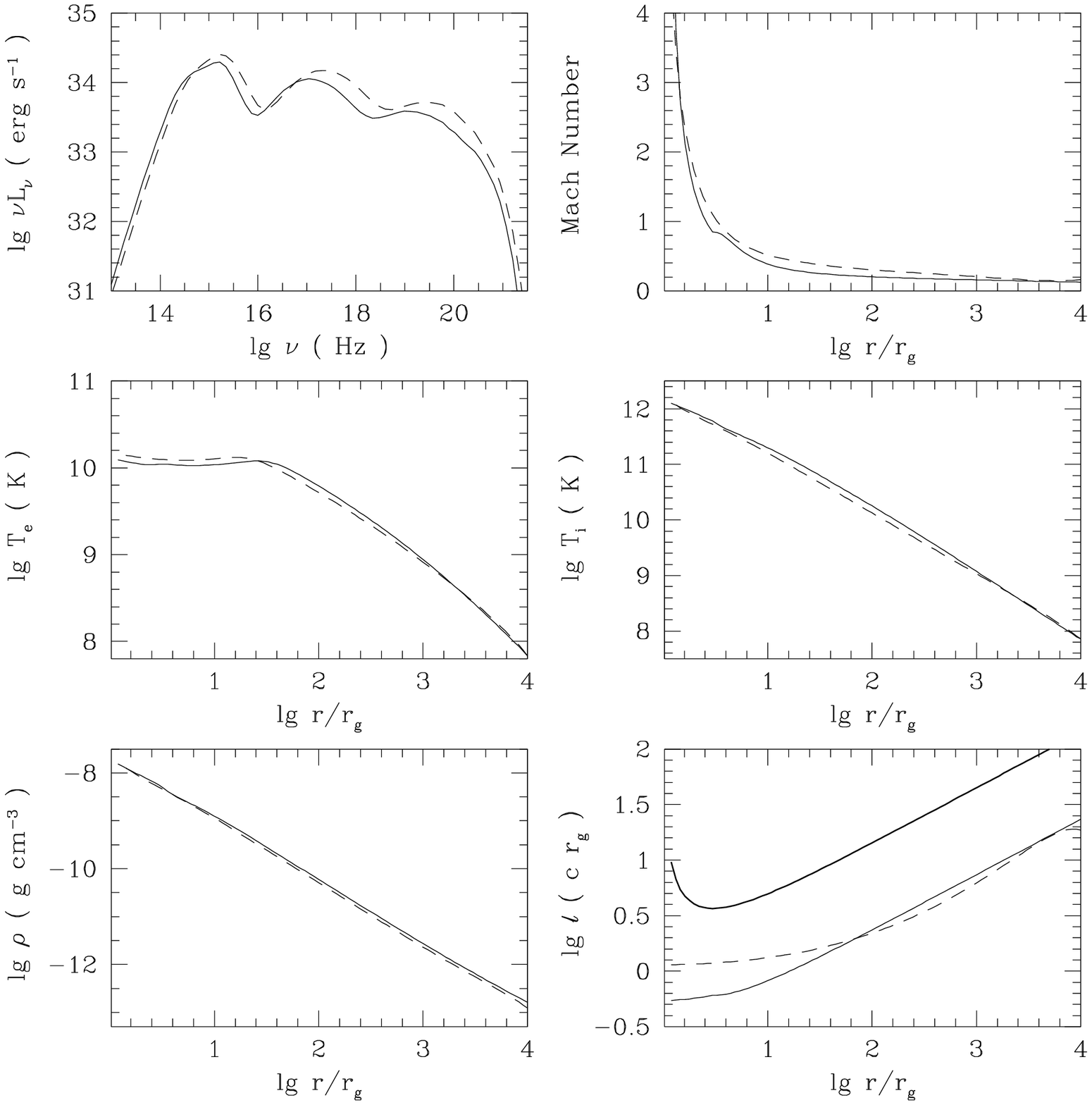} \vspace{.4in}
\caption{The simplified (solid) and exact (dashed) global ADAF solutions
for $\dot{M}= 10^{-3}\dot{M}_{\rm Edd}, ~\alpha=0.1$. The
parameters are $f_0=0.33, ~j=1.08$.}
\end{figure}

\begin{figure} \epsscale{0.9} \plotone{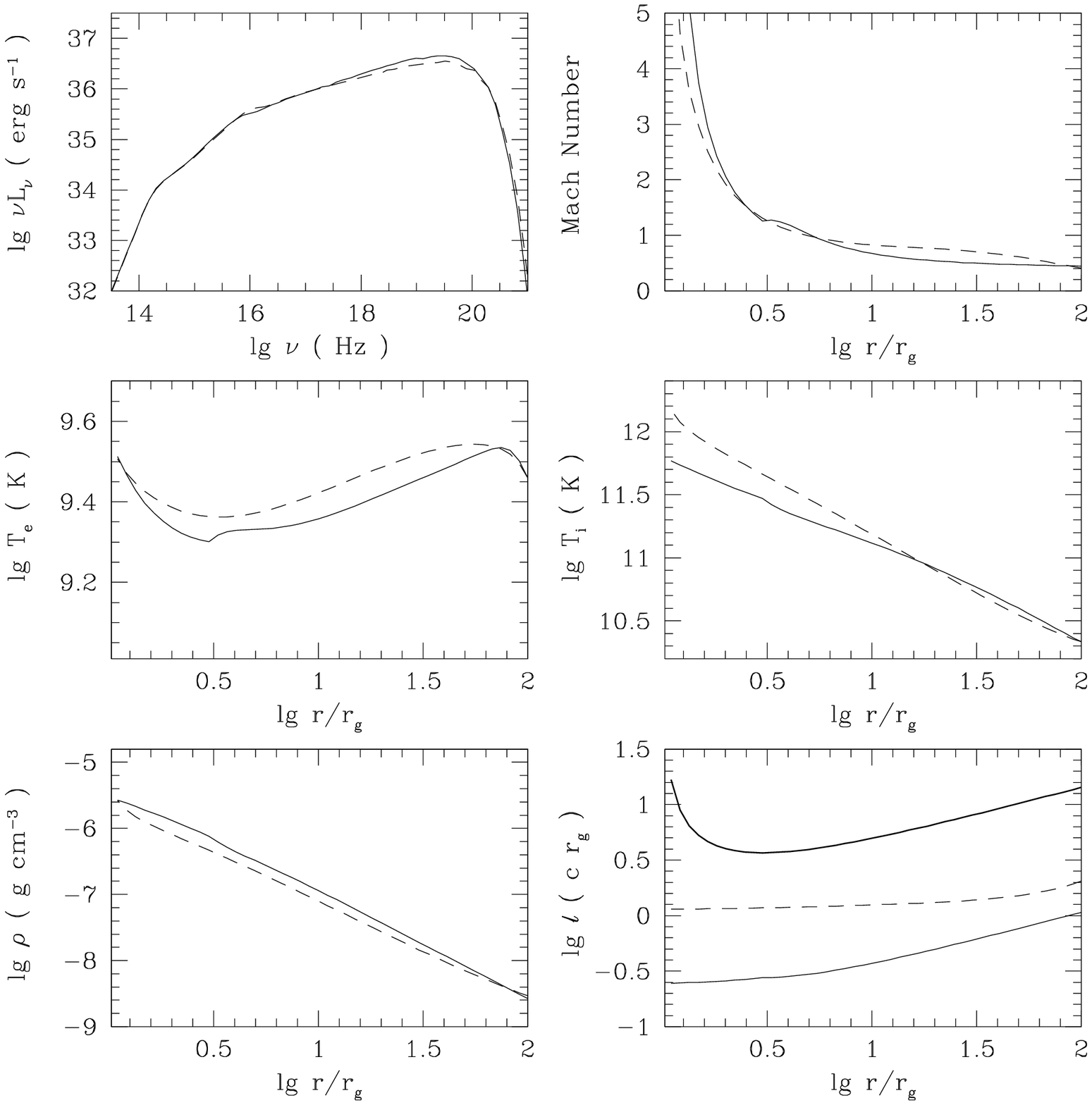} \vspace{.4in}
\caption{The simplified (solid) and exact (dashed) global ADAF solutions
for $\dot{M}= 10^{-1}\dot{M}_{\rm Edd}, ~\alpha=0.1$. The
parameters are $f_0=0.15, ~j=0.49$.}
\end{figure}

\end{document}